\documentclass[%
reprint,
amsmath,amssymb,
aps,
]{revtex4-1}

\usepackage{graphicx}
\usepackage{dcolumn}
\usepackage{bm}
\usepackage{hyperref}
\usepackage[mathlines]{lineno}
\usepackage{float}
\usepackage{color}

\begin{document}


\title{Directional detection of dark matter streams}

\author{Ciaran A. J. O'Hare} 
\email{ciaran.ohare@nottingham.ac.uk}
\author{Anne M. Green}%
\email{anne.green@nottingham.ac.uk}
\affiliation{%
School of Physics and Astronomy, University of Nottingham, University Park, Nottingham, NG7 2RD, UK
}

\date{\today}

\begin{abstract}
Directional detection of WIMPs, in which the energies and directions of the recoiling nuclei are measured, currently presents the only prospect for probing the local \textit{velocity} distribution of Galactic dark matter. We investigate the extent to which future directional detectors would be capable of probing dark matter substructure in the form of streams. We analyse the signal expected from a Sagittarius-like stream and also explore the full parameter space of stream speed, direction, dispersion and density. Using a combination of non-parametric directional statistics, a profile likelihood ratio test and Bayesian parameter inference we find that within acceptable exposure times ($\mathcal{O}(10)$ kg yr for cross sections just below the current exclusion limits) future directional detectors will be sensitive to a wide range of stream velocities and densities. We also examine and discuss the importance of the energy window of the detector.
\end{abstract}


\maketitle

\section{\label{sec:introduction}Introduction}
Attempts to directly detect weakly interacting massive particles (WIMPs) using liquid and solid-state detectors have a long history. A key goal of this field is the detection of one (or more) of the `smoking gun' signals of direct detection: annual modulation~\cite{Drukier,Freese:2012xd}, material dependence (e.g. Ref.~\cite{Peter1}) and direction dependence~\cite{Spergel}.  Much theoretical work has been carried out investigating how the WIMP particle physics (mass and cross-section) and astrophysics (local density and velocity distribution) could be inferred from the energies (e.g. Refs.~\cite{Pato:2010zk,Peter:2011eu,Peter1}) or energies and directions~\cite{Copi1,Copi2,Billard2010,Alenazi,LeePeter} of WIMP induced nuclear recoil events.

The directionality of the WIMP event rate is due to the motion of the Sun with respect to the Galactic halo. We are moving towards the constellation Cygnus  and hence the nuclear recoils are expected to be strongly peaked in the direction opposite to this. The strength of the signal is expected to be large; an anisotropic set of recoils can be discriminated from isotropic backgrounds with as few as 10 events~\cite{Copi1, Green},  while the peak recoil direction can be measured,  and the Galactic origin of the scattering particle confirmed, with around $30-50$ events~\cite{Billard2,Green2}.

Practically directional detection is still in its early stages, with a number of prototype detectors in the process of development. Directional detection is typically achieved using gas time projection chambers (TPCs) which contain low pressure gases such as CF$_4$, CS$_2$, C$_4$H$_{10}$ or $^3$He (see Ref.~\cite{Ahlen} for a review). After an interaction with a WIMP, the recoiling nucleus leaves an ionisation track in its wake. The 3-dimensional reconstruction of the recoil track then allows the full energy and direction dependence of the WIMP recoil spectrum to be measured and, in principle, the WIMP velocity distribution can be inferred~\cite{Gondolo}. The detectors currently in the prototype phase include DMTPC~\cite{DMTPC}, DRIFT~\cite{DRIFT}, MIMAC~\cite{MIMAC}, and NEWAGE~\cite{NEWAGE}. Directional detection offers several theoretical advantages over its non-directional counterpart. Firstly and foremost, there are no known backgrounds able to mimic the WIMP signal in its characteristic peak direction. Furthermore it offers the only prospect for constraining the local {\it velocity} distribution of the dark matter.

The dependence of the experimental signals on the form of the local WIMP velocity distribution has attracted a lot of attention in the literature, as there is significant uncertainty in the form of the velocity distribution, and the parameters on which it depends (for a review see Ref.~\cite{GreenRev}). Data from direct detection experiments are typically compared using the standard halo model (SHM), for which the velocity distribution is an isotropic Maxwell-Boltzmann distribution. The shape of the true local velocity distribution is expected to depart significantly from this simple model~\cite{Kuhlen, Mao, Mao2, Lisanti}. However the use of the SHM is at least somewhat justified for several reasons.  Firstly there is no fully agreed upon alternative parametrisation of the velocity distribution and secondly the expected deviations from the SHM are unlikely to affect the analysis of data from the \textit{current} generation of non-directional detectors, provided the free parameters are appropriately marginalised over~\cite{Pato:2012fw}.

Nevertheless, results from N-body and hydrodynamical simulations of galaxy formation show deviations from the putative smooth and isotropic SHM, including tidal streams~\cite{Maciejewski}, a dark disk~\cite{Read2,Read3} and debris flows~\cite{Kuhlen2,Lisanti2}, which could be detectable by future experiments. In this paper we focus on substructure in the form of tidal streams, which result from the disruption of sub-halos accreted by the Milky Way. There are hints that such a feature may pass through the Solar neighbourhood. A contiguous group of stars moving with a common velocity, that are possibly part of a tidal tail from the Sagittarius dwarf galaxy, have been observed nearby~\cite{Yanny,Majewski,Newberg,Sollima}. Moreover it has been argued that the dark matter content of the stream is likely to be significantly more extended than the stellar content and to have an offset of as much as a few kpc~\cite{Purcell}.

We examine the capabilities of future directional dark matter detectors to detect a dark matter stream. In Sec.~\ref{sec:theory}, we overview the calculation of the directional event rate, including the relevant particle physics and astrophysics input. We will begin our statistical analysis in Sec.~\ref{sec:directionalstats} with non-parametric directional tests, studying first the example of a Sagittarius-like stream and then extending the analysis to streams in general. Then in Sec.~\ref{sec:llhoodratio} we perform likelihood analyses, using Bayesian inference to reconstruct the parameters of a Sagittarius-like stream and a profile likelihood ratio test to assess the detectability of general streams in a parametric way. Finally we conclude with a discussion of our results in Sec.~\ref{sec:summary}

\section{\label{sec:theory}Directional detection}
	\subsection{Particle physics}
		The directional event rate is typically written in the form of a double differential~\cite{Spergel,Lewin,Gondolo}:
		\begin{equation}\label{dR/dEdO}
			\frac{\textrm{d}^2R}{\textrm{d}E \textrm{d}\Omega_q} = \frac{\rho_0 \sigma_N}{4\pi m_{\chi} \mu^2} F^2(E) \hat{f}_\textrm{lab}(v_{\textrm{min}}(E),\hat{\textbf{q}}) \,.
		\end{equation}
		Here, $R$ is the event rate per unit time per unit mass of detector which is expressed as a function of the direction and energy of the recoiling nucleus $\mathbf{E} = E\hat{\mathbf{q}}$ in the laboratory frame, where $\textrm{d}\Omega_q$ is the solid angle element around the recoil direction $\hat{\textbf{q}}$. The local dark matter density is denoted by $\rho_0$,  the WIMP-nucleus scattering cross section by $\sigma_N$, the WIMP mass by $m_\chi$ and  $\hat{f}_{\rm lab} (v_{\textrm{min}}(E),\hat{\textbf{q}})$ is the Radon transform of the local WIMP velocity distribution in the lab frame~\cite{Gondolo}, where 
				\begin{equation}
		v_{\textrm{min}} = \frac{\sqrt{2 m_N E}}{2 \mu} \, , 
		\end{equation}
 is the minimum WIMP speed that can scatter to create a recoil of energy $E$. The WIMP-nucleus reduced mass is  $\mu = m_\chi m_N / (m_\chi + m_N)$, where $m_N$ is the nucleus mass.

		The nuclear form factor, $F(E)$, is the Fourier transform of the nuclear density distribution. It accounts for the decrease in the effective cross section for scattering events with non-zero momentum transfer. We use the Helm form factor,
		\begin{equation}
			F^2(E) = \left[\frac{3j_1(qr_n)}{qr_n}\right]^2 e^{-q^2 s^2} \,,
		\end{equation}
		where $r_n = 1.2 A^{1/3}$ fm, where $A$ is the mass number of the nucleus and $s = 0.9$ fm~\cite{Lewin,Engel}. For light targets such as $^{19}$F, considered in this work, the form factor can be approximated as a simple exponential
		\begin{equation}\label{formfactorapprox}
			F^2(E) = e^{-\alpha(qr_n)^2} \,,
		\end{equation}
		where the value of $\alpha$ depends on the nucleus under consideration. For $^{19}$F, we find $\alpha = 1.14$. 

		The WIMP-nucleus cross section $\sigma_N$ can be written in terms of the experiment-independent WIMP-nucleon cross section $\sigma_{p,n}$ which for spin-dependent scattering is,
		\begin{equation}\label{wimpnucleoncrosssection}
			\sigma_N = \frac{4}{3} \frac{\mu_N^2}{\mu_p^2} \frac{J+1}{J} \left[\frac{a_p\langle S_p \rangle + a_n \langle S_n \rangle}{a_{p,n}} \right]^2 \sigma_{p,n} \, ,
		\end{equation}
		where $\mu_p$ is the WIMP-proton reduced mass, $J$ the total spin of the nucleus, $a_{p,n}$ the model-dependent couplings to protons and neutrons, and $\langle S_{p,n} \rangle$ are the expectation values of proton and neutron spins inside the nucleus~\cite{Jungman}.
		
	\subsection{Astrophysics}\label{sec:astrophysics}
		The velocity distribution of WIMPs in the rest frame of the laboratory is obtained through a Galilean transformation of the Galactic frame distribution, $f_{\textrm{gal}}(\mathbf{v})$, by the laboratory velocity $\mathbf{v}_\textrm{lab}$,
		\begin{equation}
			f_\textrm{lab}(\textbf{v}) = f_\textrm{gal}(\textbf{v} + \textbf{v}_\textrm{lab}) \, .
		\end{equation}
		We ignore the effects of gravitational focusing of the distribution by the Earth and Sun, which is known to be small and takes place below the threshold energies of most detectors~\cite{Lee1} and also the annual modulation due to the Earth's orbit, which is also only detectable with large numbers of events~\cite{Bozorgnia}.  The lab velocity is then the sum of the velocity of the local standard of rest (LSR) and the peculiar velocity of the Sun: $\textbf{v}_\textrm{lab} = \textbf{v}_\textrm{LSR} +  \textbf{v}_\odot$. The velocities are represented in Cartesian co-ordinates, defined using the local Galactic co-ordinate system, oriented such that the local standard of rest (LSR) points in the $y$-direction. The speed of the LSR is typically taken to be $v_\textrm{LSR} = 220$ km s$^{-1}$ although there is some uncertainty in this value \cite{Kerr,Bovy}. The most recent measurement of the peculiar velocity of the Sun is $\textbf{v}_\odot = (6.0,10.6,6.5) \pm (0.5,0.8,0.3)$ km s$^{-1}$~\cite{Bobylev}. 
		
		The SHM (which has a density profile $\rho \sim r^{-2}$, and hence a flat rotation curve) has an isotropic Maxwell-Boltzmann (MB) velocity distribution which is typically truncated by hand at the Galactic escape speed, $v_\textrm{esc}$, 
		\begin{equation}\label{SHM}	
			f^{\textrm{MB}}_{\textrm{gal}}(\textbf{v};\sigma_v,v_\textrm{esc}) = \frac{1}{N_{\textrm{esc}} (2 \pi \sigma_v^2)^{3/2}} \exp{\left(-\frac{|\textbf{v}|^2}{2 \sigma_v}\right)}\, \theta(v_\textrm{esc} - |\textbf{v}|)\, ,
		\end{equation}
		where the normalisation constant, $N_{\rm esc}$, is given by
		\begin{equation}
			N_{\textrm{esc}} = \textrm{erf}\left(\frac{v_{\textrm{esc}}}{\sqrt{2} \sigma_v}\right) - \sqrt{\frac{2}{\pi}} \frac{v_{\textrm{esc}}}{\sigma_v} \exp\left(-\frac{v_{\textrm{esc}}^2}{2 \sigma_v^2}\right) \, .
		\end{equation}
		The dispersion $\sigma_v$ is related to the speed of the LSR via  $\sigma_{v}= v_{\rm LSR}/\sqrt{2}$ and we take the Galactic escape speed to be $v_\textrm{esc} = 533 \, {\rm km \, s}^{-1}$~\cite{RAVE}. 
		
		The function $\hat{f}$ that appears in the differential recoil spectrum, Eq.~(\ref{dR/dEdO}), is the Radon transform of this distribution at $w=v_{\textrm{min}}$ which is defined as ~\cite{Radon},
		\begin{equation}
			\hat{f}(w,\hat{\textbf{q}}) = \int \delta(\textbf{v} \cdot \hat{\textbf{q}} - w) f(\textbf{v})\, \textrm{d}^3 v \,.
		\end{equation}
It is most efficient to calculate the Radon transform of the velocity distribution in the lab frame by first calculating it in the Galactic frame and then transforming to the lab frame via
		\begin{equation}
			\hat{f}_{\textrm{lab}}(v_{\textrm{min}},\hat{\textbf{q}};\sigma_v,v_\textrm{esc}) = \hat{f}_{\textrm{gal}}(v_{\textrm{min}} + \textbf{v}_\textrm{lab} \cdot \hat{\textbf{q}},\hat{\textbf{q}}) \, .
		\end{equation}
		The calculation for the SHM velocity distribution, Eq.~(\ref{SHM}), yields~\cite{Gondolo}
		\begin{eqnarray}\label{SHMradontransform}
			\hat{f}^{\textrm{MB}}_\textrm{lab}(&v_{\textrm{min}}&,\hat{\textbf{q}};\sigma_v,v_\textrm{esc}) = \frac{1}{N_{\textrm{esc}} (2 \pi \sigma_v^2)^{1/2}} \\ &\times& \left[ \exp{\left(-\frac{|v_\textrm{min}+\textbf{v}_\textrm{lab} \cdot \hat{\textbf{q}}|^2}{2 \sigma_v^2}\right)}  - \exp\left(-\frac{v_\textrm{esc}^2}{2 \sigma_v^2} \right) \right] \,. \nonumber
		\end{eqnarray}

		The simplest way to incorporate substructure into a halo model is by combining two distributions, a background smooth halo model and an additional component corresponding to the substructure. We adopt the MB velocity distribution function for the background halo model, as deviations from this are likely to be too small to have an appreciable effect on the detection of streams. We assume that a single tidal stream makes up a fraction $\xi = \rho_\textrm{str}/\rho_0$ of the local dark matter density, so that the total velocity distribution of the halo+stream model is given by
		\begin{equation}
			f^\textrm{h+s}_\textrm{gal}(\textbf{v}) = (1-\xi)f^\textrm{halo}_\textrm{gal} + \xi f^\textrm{str}_\textrm{gal} \, ,
		\end{equation}
		where,
		\begin{eqnarray}
		 f^\textrm{halo}_\textrm{gal}(\textbf{v};\sigma_\textrm{v},v_\textrm{esc}) &=& f^{\textrm{MB}}_\textrm{gal}(\textbf{v};\sigma_v,v_\textrm{esc}) \, , \\
		 f^\textrm{str}_\textrm{gal}(\textbf{v};\textbf{v}_\textrm{str},\sigma_\textrm{str},v_\textrm{esc}) &=& f^{\textrm{MB}}_\textrm{gal}(\textbf{v}-\textbf{v}_\textrm{str};\sigma_\textrm{str},v_\textrm{esc}) \,. \quad \quad
		\end{eqnarray}
		For simplicity we have taken the velocity distribution of the stream to be a MB distribution too, but with dispersion $\sigma_\textrm{str}$  and a mean velocity $\textbf{v}_\textrm{str}$ in the Galactic frame
		
\begin{figure*}
			
			\includegraphics[trim = 0mm 0 0mm 0mm, clip, width=\textwidth]{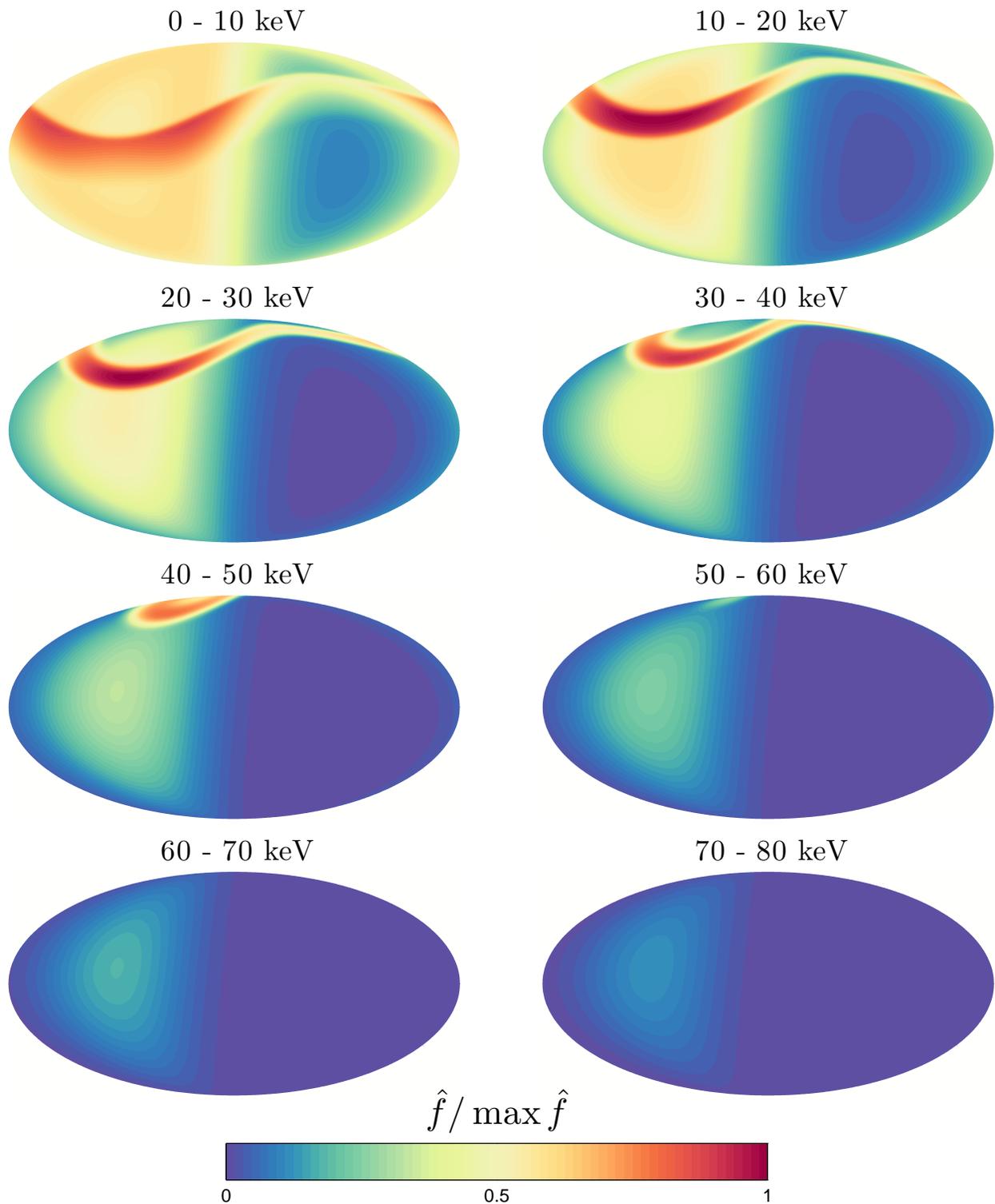}
			\caption{Mollweide projection skymaps of the rescaled Radon transform of the velocity distribution, integrated over $10$ keV energy bins, for the halo+stream model described in Sec. \ref{sec:astrophysics}, for a Sagittarius-like stream with density fraction $\xi = 0.1$ for an $m_\chi = 50$ GeV WIMP scattering off $^{19}$F.}
			\label{streamskymap}
		\end{figure*}

		The full Radon transform for the halo+stream model is therefore
		\begin{eqnarray}\label{streamradon}
			\hat{f}^\textrm{h+s}_\textrm{lab}(v_\textrm{min},\hat{\textbf{q}}) = (1-\xi)\hat{f}_\textrm{gal}^\textrm{MB}(v_\textrm{min} + \textbf{v}_\textrm{lab} \cdot \hat{\textbf{q}},\hat{\textbf{q}}; \sigma_v, v_\textrm{esc}) \nonumber \\
			+ \, \xi\hat{f}_\textrm{gal}^\textrm{MB}(v_\textrm{min} + (\textbf{v}_\textrm{lab} - \textbf{v}_\textrm{str}) \cdot \hat{\textbf{q}},\hat{\textbf{q}}; \sigma_\textrm{str}, v_\textrm{esc}) \,, \nonumber \\
			  \, 
		\end{eqnarray}
where $\hat{f}_\textrm{gal}$ is the Radon transform of the Galactic frame Maxwell-Boltzmann distribution. Figure~\ref{streamskymap} shows the angular and energy dependence of the Radon transform of the halo+stream distribution.  We have taken the stream velocity and dispersion to correspond to a Sagittarius-like stream, $\sigma_\textrm{str} = 10$ km s$^{-1}$ and $\textbf{v}_\textrm{str} = 400 \times (0,0.233,-0.970)$ km s$^{-1}$~\cite{Savage,Freesestream}, however  we have set the stream density fraction to $\xi=0.1$ in order to make the feature more prominent in the image. This is larger than the maximum stream density found in simulations~\cite{Maciejewski, Vogelsberger}.

\begin{figure}

			\includegraphics[trim = 5mm 0 0mm 0mm, clip, width=0.45\textwidth]{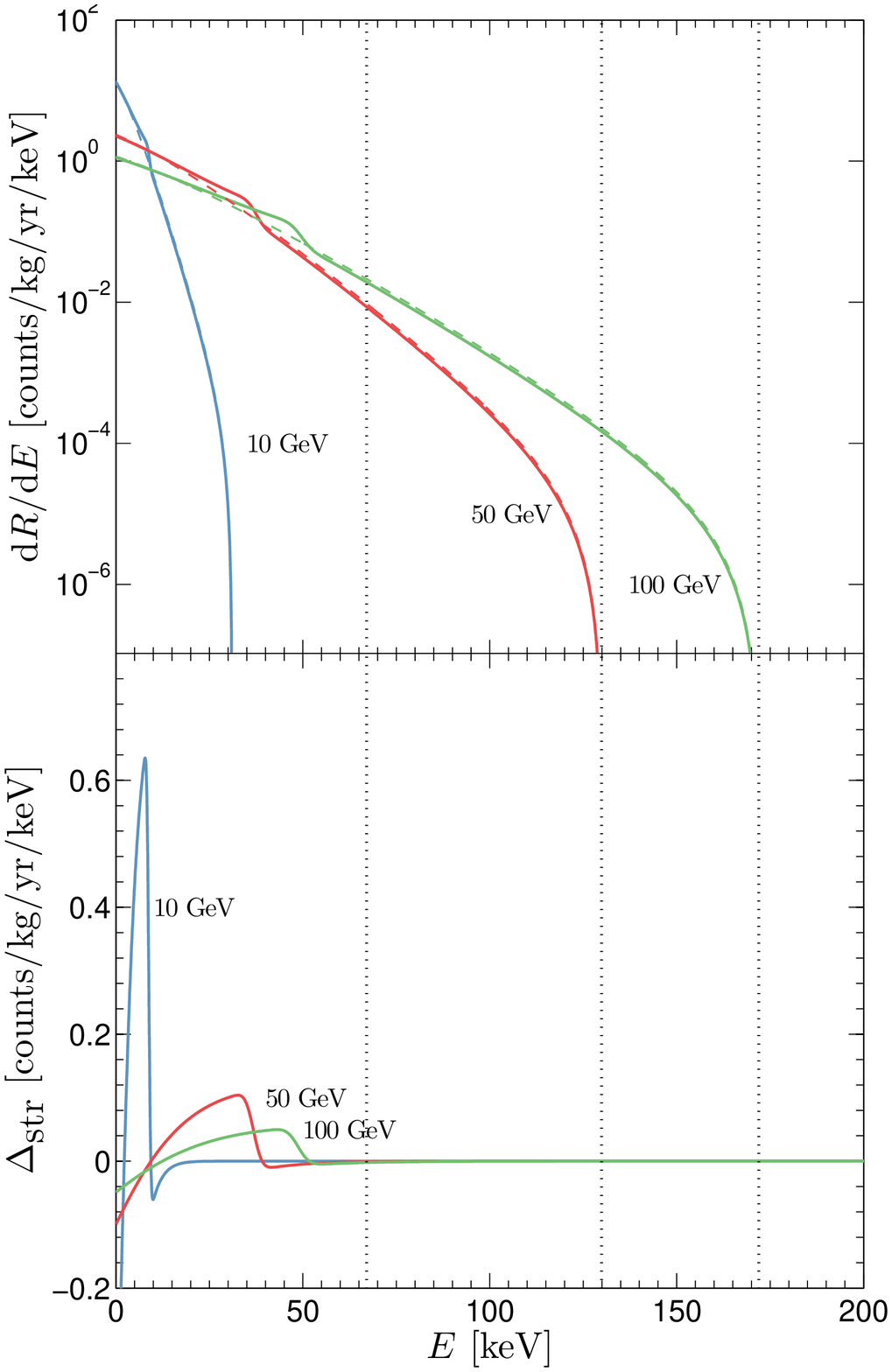}
			\caption{Top panel: Non-directional differential event rate for the halo (dashed lines) and halo+stream models (solid lines) as a function of recoil energy and $m_{\chi} =$ 20 (blue), 50 (red) and 100 (green) GeV. The dotted vertical lines indicate the cutoffs due to the escape speed. Bottom panel: difference between the event rate in the halo and halo+stream models, $\Delta_{\rm str}$, (defined in Eq. (\ref{Delta_str})) as a function of energy for the same WIMP masses. The input parameters are the same as in Fig. \ref{streamskymap}.}
			\label{energyspectrum}
		\end{figure}

		The top panel of Fig.~\ref{energyspectrum} displays more clearly the energy dependence of the signal and the dependence on the WIMP mass. The energy dependence is found by integrating Eq.~(\ref{dR/dEdO}) over angles,
		\begin{equation}
		\frac{\textrm{d}R}{\textrm{d}E} = \int \frac{\textrm{d}^2R}{\textrm{d}E \textrm{d}\Omega_q} \, \textrm{d} \Omega_q \, .
		\end{equation} 
We have fixed the values of the local density and the cross section to $\rho_{0} = 0.3$ GeV cm$^{-3}$ and $\sigma_p = 10^{-3}$ pb in this and subsequent  plots.  In the bottom panel we show the difference between the differential event rate for the halo+stream and halo-only models defined as
		\begin{equation}\label{Delta_str}
		 \Delta_{\textrm{str}} = \frac{\textrm{d}R}{\textrm{d}E}\bigg|_{\xi=0}-\frac{\textrm{d}R}{\textrm{d}E} \, .
		\end{equation}
		The stream manifests itself in the energy dependence of the event rate as a step-like feature that appears at higher energies for more massive WIMPs (this effect has been studied further in the context of annual modulation in Ref.~\cite{Freesestream}). The deviation from the smooth halo event rate is sharper and larger in size for lighter WIMPs.

	\subsection{Scattering simulation}
		To study the detectability of a stream we simulate mock experiments by Monte Carlo generating nuclear recoils. To do this we build the velocity flux distribution, sample velocities from this distribution, and scatter the resulting velocities off the chosen nucleus. We simulate a MIMAC-like experiment with target CF$_4$, in which the scattering is dominated by $^{19}$F.

		The energy of a recoil is given by,
		\begin{equation}
			E = \frac{2 \mu^2 v^2 \cos^2\theta}{m_N} \,,
		\end{equation}
		where $\theta$ is the scattering angle between the initial WIMP direction and the direction of the recoiling nucleus. The scattering angle is related to the centre of mass scattering angle, $\theta_\textrm{com}$, by
		\begin{equation}
		\theta = \frac{\pi}{2} - \frac{\theta_\textrm{com}}{2} \,.
		\end{equation}
		The scattering process is isotropic in the centre of mass (com) frame so the com angle is taken to be isotropically distributed, i.e. $\theta_\textrm{com}= \cos^{-1} (2u-1)$ where $u \in [0,1)$ is a uniformly distributed random variate. The recoil vector is generated by deflecting the initial WIMP direction by the elevation angle $\theta$ and then rotating the deflected vector by a uniformly random angle $\phi \in [0,2\pi)$ around the initial WIMP direction. The correction to the event rate due to the nuclear form factor is taken into account by calculating $F^2(E)$ for each recoil and then discarding each recoil with a probability $1-F^2(E)$. 

		Finally we must account for the exposure and background  of the detector. We split the total number of events seen into WIMP and background events,	$N_\textrm{tot} = N_\textrm{wimp}  +  N_\textrm{bg}$. The $N_\textrm{bg}$ events are generated from an isotropic distribution with a flat energy spectrum.  Ref.~\cite{Billard} showed that the background model assumed has little effect in the initial detection phase, provided the detector has good sense recognition. We show in Appendix A that this is also the case for the reconstruction of the stream parameters using the profile likelihood ratio test statistic outlined in Sec. \ref{sec:llhoodratio}.
To specify the number of events we vary the exposure, $\mathcal{E}$, measured in kg yr. This is the most physical way of adjusting the number of background events as it is independent of the WIMP mass or halo parameters. The total number of events is related to the exposure time and the total rate by
		\begin{equation}
			N_\textrm{tot} = \mathcal{E}R_\textrm{tot} = \mathcal{E} (R + R_{\rm bg}) \,,
		\end{equation}	
where $R_\textrm{bg}$ is the background event rate and $R$ is the total WIMP rate
\begin{equation}			
			R= \int_{E_\textrm{th}}^{E_\textrm{max}} \int_{\Omega_q} \frac{\textrm{d}^2 R}{\textrm{d}E \textrm{d}\Omega_q}\, \textrm{d}\Omega_q \, \textrm{d}E  \,.
		\end{equation}

	\subsection{Parameter values}
		As seen in  Eq.~(\ref{dR/dEdO}), the local WIMP density $\rho_0$ and spin-dependent (SD) WIMP-nucleon cross section $\sigma_p$ do not affect the shape of the recoil spectrum, only its amplitude, i.e. the total number of events seen in a given exposure time.  We fix these parameters to $\rho_0 = 0.3 \textrm{ GeV cm} ^{-3}$ and $\sigma_p = 10^{-3}$ pb. This value of $\sigma_p$ lies just below current exclusion limits from XENON100~\cite{Xenon}, so our results are optimistic, however they still hold if $\sigma_p$ is smaller, but for larger exposure times. 

\begin{figure}
			\centering
			\includegraphics[trim = 5mm 0mm 5mm 5mm, clip, width=0.5\textwidth]{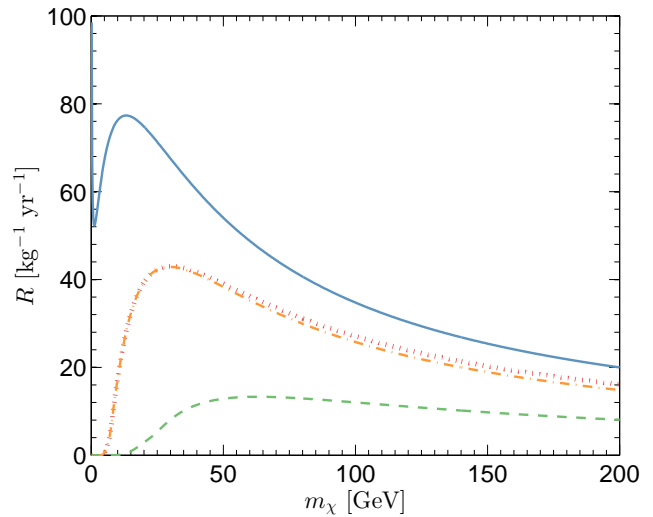}
			\caption{Total WIMP scattering rate, $R$, for a ${}^{19}$F target as a function of WIMP mass, $m_{\chi}$, for a range of experimental energy windows $[E_\textrm{th},E_\textrm{max}]$: $[0,\infty]$ keV (blue line), $[5,100]$ keV (red dotted), $[5,50]$ keV (orange dash-dot), $[20,100]$ keV (green dashed).}
			\label{R_m_chi}
		\end{figure}

		As we have displayed in Fig. ~\ref{energyspectrum} the WIMP mass affects the energy dependence of the differential event rate. However $m_\chi$ also appears in the pre-factor of Eq.~(\ref{dR/dEdO}) hence the amplitude of the signal will also depend on the WIMP mass. In Fig.~\ref{R_m_chi} we show the dependence of the signal event rate, $R$, on the WIMP mass for three different experimental energy windows, $[E_\textrm{th},E_\textrm{max}]$. We include the dependence on the energy window as this will play a crucial role in the detectability of streams. A change in the input WIMP mass will affect the results of the statistical tests due to both the change in the number of events and also the prominence of the stream in the data. For heavier WIMPs one would expect fewer total events and a stream signal that is more dispersed in energy, making it harder to detect a stream. When studying the effects of other parameters and for the most computationally demanding calculations we fix the input WIMP mass to $m_{\chi} = 50 \, {\rm GeV}$ however, where appropriate, we also consider input WIMP masses of 10 and 100 GeV.

		\begin{table}[t]
		  \begin{center}
		\begin{ruledtabular}
		    \begin{tabular}{l l l}
		    
			WIMP: 
				& $m_\chi$ & 50 GeV \\ 
				& $\sigma_p$ (SD) & 10$^{-3}$ pb \\ \hline 
			Halo: 
				& $\rho_0$ & 0.3 GeV cm$^{-3}$ \\ 
				& $\sigma_v$ & $v_\textrm{LSR}/\sqrt{2}$ \\
				& $v_\textrm{esc}$ & 533 km s$^{-1}$  \\ 
				& $\textbf{v}_\textrm{lab}$ & $(6.0,230.6,6.5)$ km s$^{-1}$ \\  \hline	
			Experiment: 
				& $m_N$ & 18.998 amu (F) \\
				& $E_\textrm{th}$ & 5, 20 keV \\
				& $E_\textrm{max}$ & 50, 100 keV \\
				& $R_\textrm{bg}$ & 10 kg$^{-1}$ yr$^{-1}$ \\ \hline
			 Sgt. stream:
				& $\textbf{v}_\textrm{str}$ & $400 \times (0,0.233,-0.970)$ km s$^{-1}$ \\
				& $\sigma_\textrm{str}$ & 10 km s$^{-1}$ \\
				& $\xi_\textrm{str}$ & 0.1 \\
		    \end{tabular}
		\end{ruledtabular}
		  \end{center}
		  \caption{Benchmark WIMP, halo, experimental and stream parameters used in Sec.~\ref{sec:directionalstats}.}
		\label{tab:partable}
		\end{table}

		In the following sections we will compare the power of several different statistical tests using the benchmark WIMP, halo, experimental and stream parameter values listed in Tab. \ref{tab:partable}. For the other background halo parameters the values used are taken from the literature (as outlined in Sec.~\ref{sec:astrophysics}) but the precise values are not expected to impact the results significantly. The parameters used to describe the stream are the stream density fraction, $\xi$, velocity, $\textbf{v}_\textrm{str}$, and velocity dispersion, $\sigma_\textrm{str}$. When describing the properties of the stream we will often parameterise the direction in which it is moving in terms of the angle between $\textbf{v}_\textrm{str}$ and $\textbf{v}_\textrm{lab}$,
		\begin{equation}\label{deltatheta}
		      \Delta \theta = \cos^{-1}(\hat{\textbf{v}}_\textrm{lab} \cdot \hat{\textbf{v}}_\textrm{str}) \,.
		\end{equation}
		From symmetry considerations the performance of the statistical tests is only sensitive to this quantity. For the stream density and dispersion, unless otherwise stated, we will use $\xi = 0.1$ and $\sigma_\textrm{str} = 10$ km s$^{-1}$.

The experimental energy window, $[5,50]$ keV, reflects a plausible forecast for a CF$_4$ experiment such as MIMAC \cite{MIMAC}. 
As we will subsequently see, the threshold energy, $E_{\rm th}$, and the maximum energy, $E_\textrm{max}$, play an important role in determining the detectability of streams. 
Therefore we also study the effects of (optimistically) raising $E_\textrm{max}$ to 100 keV and (less optimistically) raising $E_\textrm{th}$ to 20 keV. Lastly we assume that the scattering is only off $^{19}$F with pure proton coupling, so in Eq.~(\ref{wimpnucleoncrosssection}) we take $J = 1/2$, $a_p = 1$, $a_n = 0$ and $\langle S_p \rangle = 0.5$.

\section{Directional statistics}\label{sec:directionalstats}
There are two steps in characterising a stream. Firstly the stream must be detected, and then its parameters (e.g. density and velocity) can be measured. First we consider non-parametric statistical tests. These tests use the direction information only and have been used in the past to determine the number of events required to distinguish a WIMP signal from isotropic backgrounds~\cite{Morgan, Green}. The advantage of a non-parametric analysis, is that it is not necessary to assume a particular model for the smooth component of the halo, and the results are valid provided the basic hypotheses that define the statistical tests are satisfied. However a notable disadvantage of non-parametric analysis is that it is less powerful than parametric analysis (in the sense that, for a given data set, the detection significance will be lower). Therefore we will subsequently perform a parametric analysis using a likelihood in Sec. \ref{sec:llhoodratio}

\begin{figure*}
			\centering
			\includegraphics[trim = 50mm 0 30mm 0mm, clip, width=\textwidth]{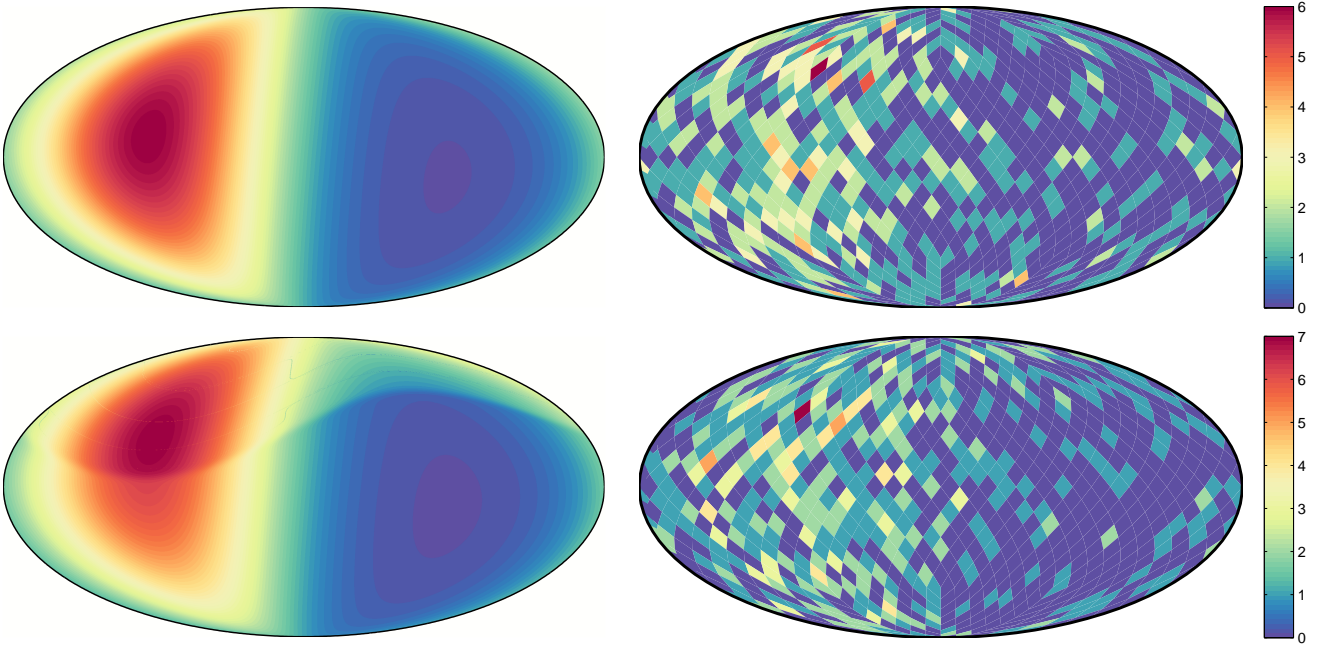}
			\caption{Halo (top row) and halo+stream (bottom row) directional signals for an idealised detector with an infinite exposure (left, normalised to unity) and for a realistic detector (right, showing event counts) with finite exposure, angular resolution and experimental backgrounds present. The signals have been generated using the parameters listed in Tab. \ref{tab:partable} with an energy window of $[5,100]$ keV.}
			\label{signal_comparison_skymap}
		\end{figure*}

The question we wish to answer is: what exposure is needed to reject, at a given confidence level, a smooth and isotropic Galactic WIMP distribution in the event that the true distribution contains an additional steam component? Figure ~\ref{signal_comparison_skymap} illustrates the problem; how does one extract information from a small quantity of spherical data? The left hand panels show the normalised directional signal observed in a perfect detector with infinite exposure, for the halo and stream parameters given in Tab. \ref{tab:partable}. The top row shows the signal with no stream present, and the bottom with the inclusion of a stream. The right hand panels show the signal expected in a real detector with an exposure of $\mathcal{E}=10$ kg yr including events from experimental backgrounds with a total rate of $R_\textrm{bg}= 10 \, {\rm kg}^{-1} {\rm yr}^{-1}$ distributed isotropically. Since the directional statistics only use information about the directions of the nuclear recoils, and not their energies, their performance is independent of the background energy spectrum. The data consist of 304 signal events in the halo-only case and 316 in the halo+stream case. The data have been binned on a sphere using a HEALPix \cite{HEALPix} equal angular area discretisation with $N_\textrm{pix} = 768$. While detecting the isotropy of a WIMP signal requires few events, finding deviations from a smooth halo will require a great deal more~\cite{Morgan}.

We construct test statistics, $\mathcal{T}>0$, that are distributed as $p_0(\mathcal{T})$ under some null hypothesis that is satisfied in the presence of a distribution without a stream and $p_1(\mathcal{T})$ when a stream is present in the data. These distributions may be known analytically for some test statistics in the null case, but if not they can be Monte Carlo generated. For a particular measured value of the test statistic, $\mathcal{T}_\textrm{obs}$, the significance is the probability of measuring $\mathcal{T}<\mathcal{T}_\textrm{obs}$ if the null hypothesis is true
		\begin{equation}\label{significance}
			S = \int_{0}^{\mathcal{T}_\textrm{obs}} p_0(\mathcal{T}) \textrm{d}\mathcal{T} \,.
		\end{equation}   
The statistical power is the probability of rejecting the null hypothesis if the null hypothesis is false. In other words it is the probability of measuring $\mathcal{T}>\mathcal{T}_\textrm{obs}$ when $\mathcal{T}$ is distributed according to $p_1(\mathcal{T})$,
		\begin{equation}
			\mathcal{P} = \int_{\mathcal{T}_\textrm{obs}}^{\infty} p_1(\mathcal{T})\textrm{d}\mathcal{T} \, .
		\end{equation}
		For detection of a stream we require $\mathcal{P}=0.95$ and define $S_{95}$ as the detection significance achievable in 95\% of experiments. 

	\subsection{Median direction}
		The median direction, $\hat{\textbf{x}}_\textrm{med}$, of a set of $N$ directions, $\lbrace \hat{\textbf{x}}_i,\, ...\, , \hat{\textbf{x}}_N\rbrace$,  is found by minimising the quantity \cite{Fisher},
		\begin{equation}
			M = \sum_{i=1}^{N} \cos^{-1}(\hat{\textbf{x}}_\textrm{med} \cdot \hat{\textbf{x}}_i) \, .
		\end{equation}
		The median recoil direction of the smooth halo distribution is $-\hat{\textbf{x}}_\textrm{lab}$, i.e. the inverse of the direction of Solar motion.
		To test whether recoils are consistent with a hypothesised median direction, $\hat{\textbf{x}}_0$, we use the $\chi^2$ test statistic which is calculated as follows~\cite{Fisher}.  First the recoil vectors $\hat{\textbf{x}}_i$ are rotated so that they are measured relative to a north pole at the sample median given by $(\theta_\textrm{med},\phi_\textrm{med}$), this is done using the rotation,
		\begin{equation}\label{RotNP}
			\hat{\textbf{x}}_i' = \textsf{R}_y(\pi/2 - \theta_\textrm{med}) \textsf{R}_z(-\phi_\textrm{med}) \hat{\textbf{x}}_i \, ,
		\end{equation}
		where $\textsf{R}_y$ and $\textsf{R}_z$ are the Cartesian rotation matrices for rotations around the $y$ and $z$ axes. After the recoil vectors have been rotated, the azimuthal angles $\phi_i'$ are then measured in this new co-ordinate system. Then the matrix,
		\begin{equation}
			\Sigma = \frac{1}{2}
			\begin{pmatrix}
				\sigma_{11} & \sigma_{12} \\
				\sigma_{21} & \sigma_{22}
			\end{pmatrix}\, ,
		\end{equation}
		where,
		\begin{eqnarray}
			\sigma_{11} &=& 1 + \frac{1}{N}\sum_{i=1}^{N} \cos 2\phi_i' \,, \\
			\sigma_{22} &=& 1 - \frac{1}{N}\sum_{i=1}^{N} \cos 2\phi_i' \,, \\
			\sigma_{12} &=& \sigma_{21} = \frac{1}{N}\sum_{i=1}^{N} \sin \phi_i' \,,
		\end{eqnarray}
		is constructed.
		Next, the recoil vectors are rotated again using Eq.~(\ref{RotNP}) but now so that they are measured relative to a north pole at the hypothesised median direction $(\theta_0,\phi_0)$. Finally the test statistic is calculated as,
		\begin{equation}
			\chi^2 = U^T \Sigma^{-1} U \, ,
		\end{equation}
		where $U$ is defined as 
		\begin{equation}
			U = \frac{1}{\sqrt{N}} 
			\begin{pmatrix}
				\sum \cos \phi_i^0 \\
				\sum \sin \phi_i^0
			\end{pmatrix} \,.
					\end{equation}
If the hypothesised median direction is correct and the number of events is $N>25$, then the test statistic is distributed according to a $\chi^2_2$ distribution. The statistical significance of a particular observed value $\chi^2_\textrm{obs}$ is then equal to the cumulative distribution function for $\chi^2_2$ at $\chi^2_\textrm{obs}$ according to Eq.~(\ref{significance}).

	\subsection{Modified Kuiper test}
		The modified Kuiper test is a test for rotational symmetry around some hypothesised direction $\hat{\textbf{x}}_0$. The test is performed by first rotating all recoil direction vectors $\hat{\textbf{x}}_i$ so that their spherical angles $(\theta_i,\phi_i)$ are measured relative to a north pole with angles $(\theta_0,\phi_0)$ prior to rotation, using the same procedure as outlined in the previous section in Eq.~(\ref{RotNP}). After the recoil vectors have been rotated, the azimuthal angles $\phi_i$ are then measured in this new co-ordinate system in units of $2\pi$, reorganised in ascending order and denoted $X_i$ such that they define a cumulative distribution $F(X)$. In the case that the data possess rotational symmetry the cumulative distribution follows $F(X) = X$. The modified Kuiper statistic quantifies deviations from this and is defined as,
		\begin{equation}
		\label{modkuiper}
			\mathcal{V}^\star = \mathcal{V} \, \left(N^{1/2} + 0.155 + \frac{0.24}{N^{1/2}}\right) \,,
		\end{equation}
		where $\mathcal{V} = D^{+} + D^{-}$
		is the unmodified Kuiper statistic and
		\begin{eqnarray}
			D^{+} &=& \textrm{max}\left(\frac{i}{N} - X_i\right) \,, \\
			D^{+} &=& \textrm{max}\left(X_i - \frac{i-1}{N}\right) \, ,
		\end{eqnarray}
		with $i = 1, ..., N$. The modification factor in Eq.~(\ref{modkuiper}) allows the distribution of the Kuiper statistic to be independent of the sample size for $N\geq8$. There is no analytic form for this distribution in the null case but there are published critical values \cite{Fisher} or the distribution can be Monte Carlo generated using any set of vectors with rotational symmetry satisfied, for example the isotropic experimental background recoils.
		\begin{figure}[h]
			\centering
			\includegraphics[trim = 0mm 0mm 0mm 0mm, clip, width=0.45\textwidth]{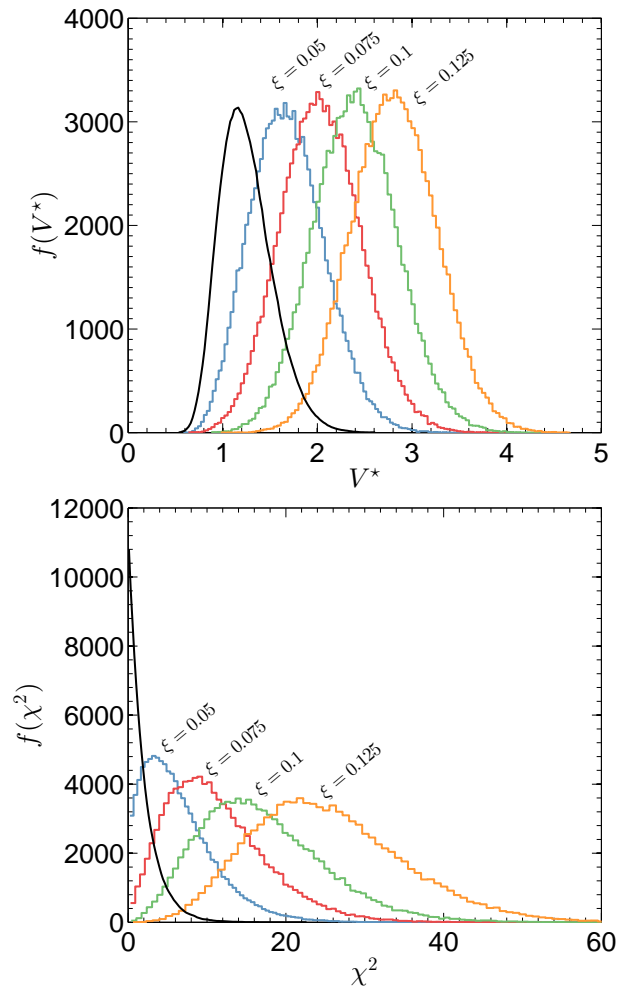}
			\caption{The distributions of the Kuiper statistic, $f(V^\star)$, (top panel) and Median direction $\chi^2$ statistic, $f(\chi^2)$, (bottom panel) built from $10^5$ Monte Carlo experiments. The distributions were generated for isotropic background recoils (black curve) and with a Sagittarius-like stream with variable density fraction $\xi$ present (coloured histograms), with increasing $\xi$ going from left to right. The stream velocity and velocity dispersion, as well as the parameters for the MB distribution of the smooth halo are given in Tab.~\ref{tab:partable}.}
			\label{dirstatsdists}
		\end{figure}

	\subsection{Results}
 Figure \ref{dirstatsdists} shows the distributions of the two test statistics when the null hypothesis is true (generated with isotropic backgrounds) and when performed on data containing recoils from a Sagittarius-like stream with varying density fraction, $\xi$. The distributions in each case were generated from $10^5$ mock experiments with an exposure of 10 kg yr. As $\xi$ is increased the degree of rotational asymmetry is increased, the median direction becomes more displaced from $-\hat{\textbf{v}}_\textrm{lab}$ and the observed distributions of the test statistics become further separated from the null distributions.

		\subsubsection{Sagittarius stream}
			\begin{figure*}
				\centering
				\includegraphics[trim = 0mm 0mm 0mm 0mm, clip, width=\textwidth]{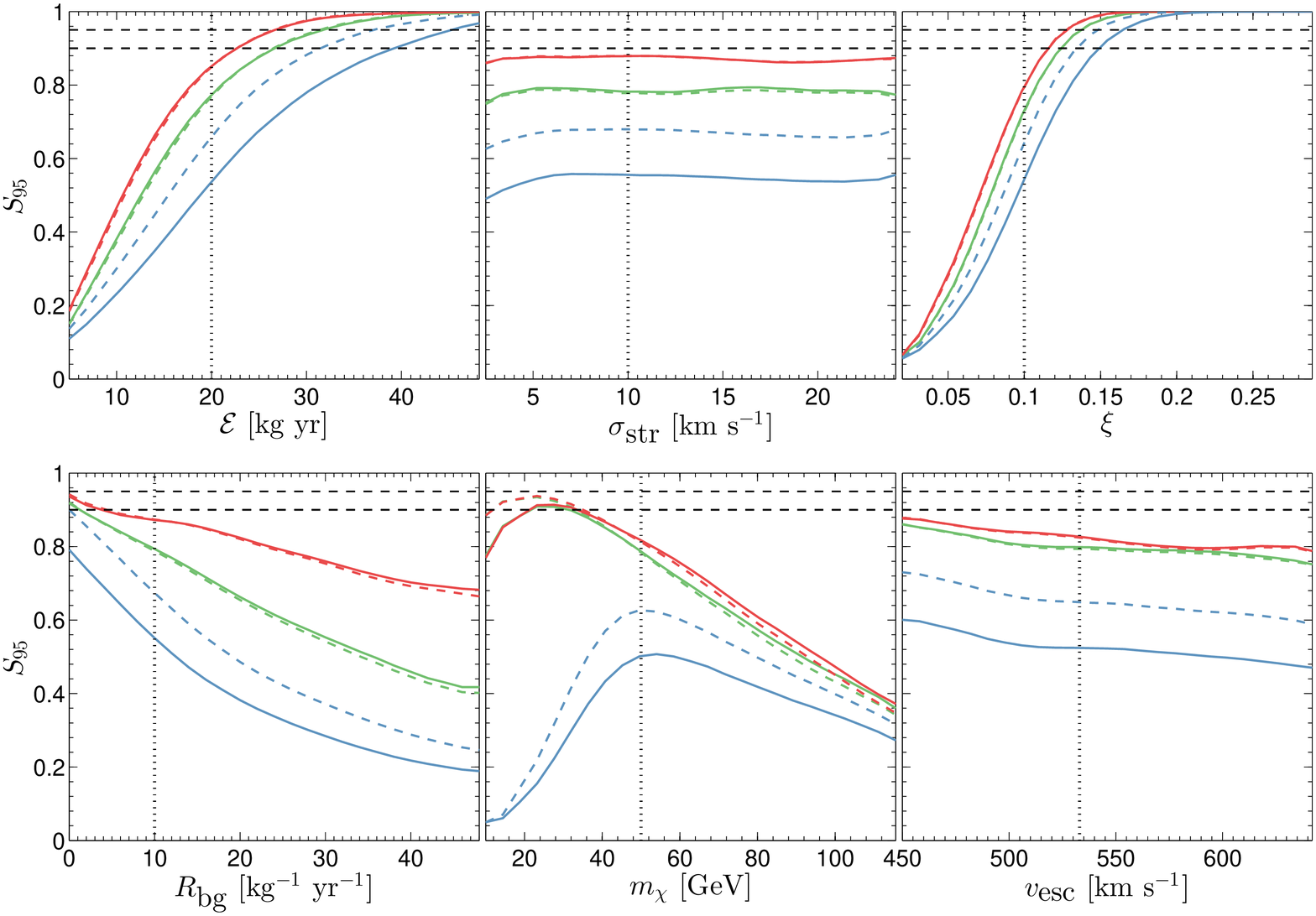}
				\caption{Significance obtainable by 95\% of experiments, $S_{95}$, as a function of exposure time, $\mathcal{E}$, stream dispersion, $\sigma_{\rm str}$, stream density fraction, $\xi$, background rate, $R_{\rm bg}$, WIMP mass, $m_\chi$, and escape speed $v_\textrm{esc}$, for the Median direction (solid lines) and Kuiper (dashed lines) tests for the Sagittarius-like stream. The results are shown for energy windows of $[5,100]$ keV (green), $[5,50]$ keV (red) and $[20,100]$ keV (blue). The horizontal dashed lines indicate the desired $0.9 - 0.95$ significance levels and the vertical dotted lines in each plot indicate the parameter values used in neighbouring plots.}
				\label{testcompare_sagit}
			\end{figure*}
			To demonstrate the performance of the tests as a function of parameters other than the stream velocity we will take as a concrete example, a Sagittarius-like tidal stream. We will assume a velocity of $\textbf{v}_\textrm{str} = 400 \times (0,0.233,-0.970)$ \cite{Savage,Freesestream} for the stream. There is sizable uncertainty in both the speed and direction of the velocity (and even whether or not dark matter from this stream passes through the Solar neighbourhood) therefore in the subsequent sub-sections we will vary the stream velocity.

			In Fig. \ref{testcompare_sagit} we plot $S_{95}$, the value of the significance $S$ given a required power $\mathcal{P} = 0.95$ (i.e. the value of significance obtainable by 95\% of hypothetical experiments) for the Kuiper and Median direction $\chi^2$ tests. We show how this quantity varies with exposure time, stream dispersion, stream density, experimental background rate, WIMP mass and escape speed. We also show the dependence of the results on the energy window of the detector, considering three examples, $[5,100]$ keV, $[5,50]$ keV and $[20,100]$ keV. The benchmark experimental and halo parameter values that are fixed when the other parameters are varied are indicated in the Figure by vertical lines.

			The results are intuitive, with longer exposure times or for streams which make up a larger fraction of the local WIMP density, the signal is stronger and hence the tests perform better. The number of stream WIMPs scales linearly with exposure time and stream density fraction hence the significance of the result scales as roughly the square root of those quantities. To detect a Sagittarius-like stream at 90 - 95\% confidence in 95\% of experiments, one would need exposures between 10 - 20 kg yr for stream densities around $\xi \sim 0.1$. 
			
			Experiments with larger background rates perform predictably poorly compared to those with fewer backgrounds to contaminate the WIMP signal. Note that the quoted value of $R_\textrm{bg}$ is the rate observed in the energy window $[5,100]$ keV and the value for the other ranges has been scaled appropriately to account for the smaller sensitivity window. The dispersion of the stream has no effect on the performance of the test, because, for dispersions of at most tens of ${\rm km} \, {\rm s}^{-1}$, the WIMPs scatter into the same angular area independent of the dispersion. The significance achieved decreases weakly as the escape speed is increased. This is because increasing the escape speed doesn't affect the recoils from stream WIMPs, but slightly increases the number of recoils from the smooth halo.
				    
			The tests perform more or less equally well, however there is a notable difference between the performance of the tests when the threshold energy is 20 keV. The lowest energy recoils are those which scatter through the largest angles. Removing the low energy recoils from the signal by increasing the threshold energy thus removes the stream WIMPs scattering furthest from the peak stream direction. As can be seen in Fig.~\ref{testcompare_sagit}, the performance of the Median direction test is weakened compared to the Kuiper test under the removal of these low energy recoils.
			
			The performance of the tests also depends on the mass of the WIMP, due mainly to the variation in the total WIMP scattering rate. As shown in Fig. \ref{R_m_chi}, typically higher mass WIMPs yield fewer events (assuming fixed exposure and local density $\rho_0$) and hence the significance achieved by the tests decreases with increasing WIMP mass. However for light WIMPs the majority of the recoils from stream WIMPs are below threshold and the significance achieved decreases. This occurs for $m_{\chi} \lesssim 50 \, (20)$ GeV for a 20 (5) keV energy threshold. As we will see in Fig.~\ref{N_wimp_str_location_cont} in Sec. \ref{sec:llhoodratio}, the number of recoils from stream WIMPs also depends on the stream velocity.

		\subsubsection{Stream velocity}
			\begin{figure}[h]
				\centering
				\includegraphics[trim = 0mm 0mm 0mm 0mm, clip, width=0.4\textwidth]{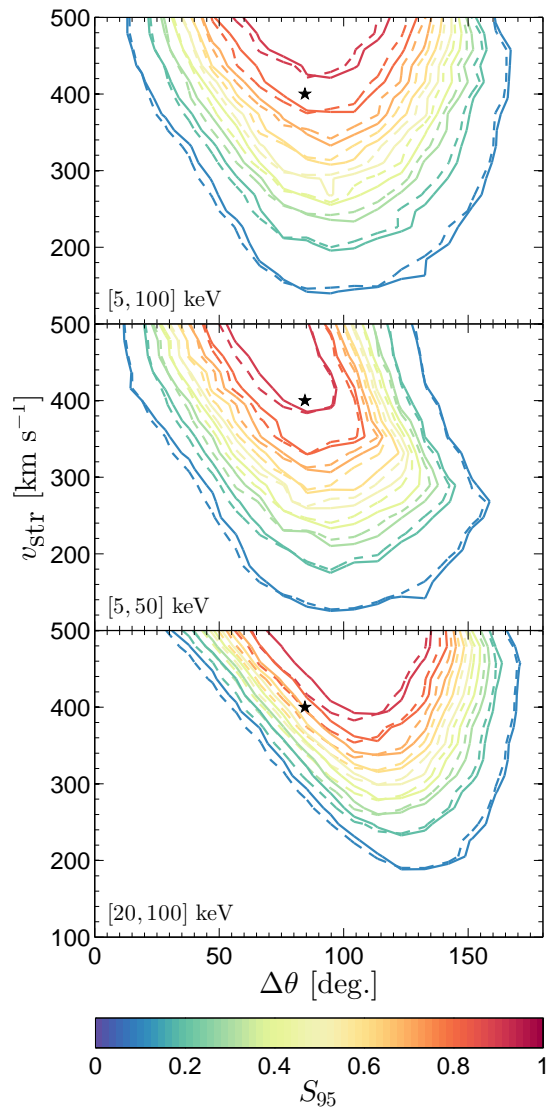}
				\caption{Significance obtainable by 95\% of experiments with a 10 kg yr exposure as a function of stream speed, $v_{\rm str}$, and the angle between the stream velocity and the direction of Solar motion, $\Delta \theta$, for a $m_{\chi} = 50 \, {\rm GeV}$ WIMP. The results for the Kuiper test are shown by the dashed lines and the Median direction $\chi^2$ test by the solid lines. The three panels are for energy windows $[5,100]$ keV, $[5,50]$ keV and $[20,100]$ keV, from top to bottom. The star denotes the speed and direction of the Sagittarius stream.}
				\label{testcomparison_DeltaTheta}
			\end{figure}

		It is  possible that a DM stream without a stellar component could pass through the 
Solar neighbourhood. Given this, and also the uncertainties in the velocity of the Sagittatius stream, we now examine how the test statistics perform when the stream velocity is varied. As mentioned previously, the stream velocity can be described by two parameters, the speed of the stream, $v_\textrm{str}$, and the angle between the lab and stream velocities, $\Delta \theta$, defined in Eq.~(\ref{deltatheta}). 

			Figure \ref{testcomparison_DeltaTheta} shows how the significance varies with the stream direction and speed, and also  the energy window of the detector. The tests do not perform equally for all stream directions. In particular the tests return a low significance when the stream is anti-aligned with the Solar velocity, as in this case the null hypotheses of rotational symmetry and inverse-Solar median direction are in fact correct. The number of events from stream WIMPs plays an important role in determining the significance which can be achieved, and for streams with small $\Delta\theta$ this number is very low and for low  streams speeds or high threshold energies it can even be zero. The symmetry in the plots is due to this dependence on both the sample size and the positioning of the stream recoils with respect to the smooth halo recoils. For the fastest streams, increasing the threshold energy results in a higher significance, this is essentially a background rejection effect;  it preferentially removes low energy smooth halo recoils and the stream is more prominent,  even with fewer overall events. The drop-off at large $\Delta\theta$ for the  [5, 50] keV window is due to the stream being boosted so that a portion of the recoils fall above the energy window. This is the reverse effect to that at low values of $\Delta\theta$ in the [20, 100] keV window, in this case the stream recoils lie below the threshold energy.

			The Kuiper and Median direction tests studied in this section only use the direction of the recoils and not their energies.  As has been shown in Ref.~\cite{LeePeter}, use of the full energy and direction data allows better constraints to be placed on the WIMP speed distribution. It may be possible to improve the tests we have studied by incorporating the energy dependence of the signal by performing the test successively on energy ordered recoils or by binning the recoils in energy. In the case where there is no stream present the hypotheses of rotational symmetry and inverse-Solar median direction are satisfied for all energies. However with a stream in the signal there will be a range of energies where the hypothesis is false. The degree to which the test is failed will initially increase with energy and then above a certain value, determined by the stream speed, the test would return a value closer to the null case. Accounting for this effect in the test statistic would decrease the overlap between the null and alternative distributions and hence increase the significance of a particular result. However it is likely to be a small effect, in Fig.~\ref{energyspectrum} one can see that the energy dependent effect of a stream is quite small and highly dependent on mass. Furthermore splitting the recoils in energy would result in a smaller number of events, and hence a loss of information, for each individual evaluation of the test statistic. Therefore it is likely that such a modified energy dependent non-parametric test would perform no better than the standard direction-only non-parametric tests we have studied, given the low numbers of recoils expected.

\section{\label{sec:llhoodratio}Likelihood analysis}
	\subsection{Likelihood function}
		We now turn our attention towards statistical tests capable of placing constraints on the properties of substructure present in the local dark matter velocity distribution as well as distinguishing between the halo and halo+stream models. For placing constraints on the parameters of the model we use Bayesian inference, whereas the test that is most appropriate for model comparison is a profile likelihood ratio test, as it uses the assumption that the null hypothesis is obtained by applying a constraint to the alternative hypothesis. The likelihood ratio test has the added advantage of being able to conveniently account for astrophysical uncertainties by treating them as nuisance parameters~\cite{Billard}. 
		
		The model we describe has 11 free parameters, $\quad \quad\boldsymbol{\theta} = \lbrace m_\chi, \sigma_p, \rho_0, \sigma_v,v_\textrm{esc},\sigma_\textrm{str},v_\textrm{str,1},v_\textrm{str,2},v_\textrm{str,3},\xi,R_\textrm{bg}\rbrace$, where we split the velocity of the stream into its three Galactic co-ordinate components. For simplicity we will again assume the smooth component of the halo has a MB distribution, but in principle the approach could be used with other velocity distributions e.g. the form of Mao et al.~\cite{Mao}.

		We define the likelihood function as the product of the probabilities for obtaining recoils with energies and directions $(E_i,\hat{\textbf{q}}_i)$ ($i = 1, ..., N_\textrm{o}$) multiplied by a Poisson factor accounting for the probability of obtaining $N_\textrm{o}$ observed events given an expected number $N_\textrm{e}(\boldsymbol{\theta})$:
		\begin{eqnarray}
			\mathcal{L}(\boldsymbol{\theta})&=& \frac{N_\textrm{e}(\boldsymbol{\theta})^{N_\textrm{o}}}{N_\textrm{o} !} e^{-N_\textrm{e}(\boldsymbol{\theta})} \nonumber \\  &\times&\prod_{i=1}^{N_\textrm{o}} \bigg[\lambda P_\textrm{wimp}(E_i,\hat{\textbf{q}}_i ; \boldsymbol{\theta}) + (1-\lambda)P_\textrm{bg}\bigg] \,, 
		\end{eqnarray}
		The expected number of events is a function of the parameters $\boldsymbol{\theta}$ and is equal to the sum of the expected number of WIMP and background events,  $N_\textrm{e}^{\textrm{wimp}}$ and  $N_\textrm{e}^{\textrm{bg}}$ respectively, 
		\begin{eqnarray}
			N_\textrm{e}(\boldsymbol{\theta}) &=& N_\textrm{e}^{\textrm{wimp}} + N_\textrm{e}^{\textrm{bg}} \,, \\ &=& \mathcal{E} \left[ \int_{E_\textrm{th}}^{E_\textrm{max}} \int_{\Omega_q} \frac{\textrm{d}^2R}{\textrm{d}E \textrm{d}\Omega_q}\bigg|_{\boldsymbol{\theta}}\, \textrm{d}\Omega_q \, \textrm{d}E+ R_\textrm{bg}\right].
		\end{eqnarray}
		The probabilities $P_\textrm{wimp}$ and $P_\textrm{bg}$ are the probabilities for an event to occur at $(E_i,\hat{\textbf{q}}_i)$ in the signal and background (no WIMP) cases respectively, i.e.
		\begin{equation}
			P_\textrm{wimp}(E_i,\hat{\textbf{q}}_i; \boldsymbol{\theta}) = \frac{1}{R} \frac{\textrm{d}^2 R}{\textrm{d}E\textrm{d}\Omega_q}\bigg|_{E_i,\hat{\textbf{q}}_i; \boldsymbol{\theta}} \,, 
				\end{equation}
		and, assuming a flat background energy spectrum,
		\begin{equation}			
			P_\textrm{bg} = \frac{1}{4\pi(E_\textrm{max}-E_\textrm{th})} \,,
		\end{equation}		
		and  $\lambda$ is the signal fraction,
		\begin{equation}
		 \lambda = \frac{R}{R+R_\textrm{bg}} \,.
		\end{equation}

		It will be difficult to constrain the escape speed, $v_\textrm{esc}$, as it only affects the ${}^{19}$F recoil spectrum at energies beyond the maximum energies we are considering. This can be overcome, as in Ref. \cite{Billard}, by treating $v_\textrm{esc}$ as a nuisance parameter and accounting for its uncertainty by hand by including an additional multiplicative term to the likelihood in the form of a Gaussian with mean and standard deviation of $v_\textrm{esc} = 533\, \pm \, 54$ km s$^{-1}$ \cite{RAVE, Piffl}.

		The WIMP density, $\rho_0$, and cross section, $\sigma_p$, are also difficult to constrain as they only appear in the amplitude of the recoil spectrum and hence only affect the number of events seen for a given exposure. Even for the quite large stream densities that we are considering here the difference between the number of events in the stream case and in the null case is small. Moreover, the two parameters are degenerate with one another meaning there is no single set of values for $\rho_0$ and $\sigma_p$ that maximise the likelihood function. Therefore we also employ a Gaussian contribution to the likelihood function for $\rho_0$ with mean and standard deviation  $\rho_0 = 0.3 \, \pm \, 0.1$ GeV cm$^{-3}$~\cite{Read}.

	\subsection{Parameter reconstruction}
	\label{param-reconstruct}
		\begin{figure*}
			\centering
			\includegraphics[trim = 0mm 0mm 0mm 0mm, clip, width=1.3\textwidth, angle=90]{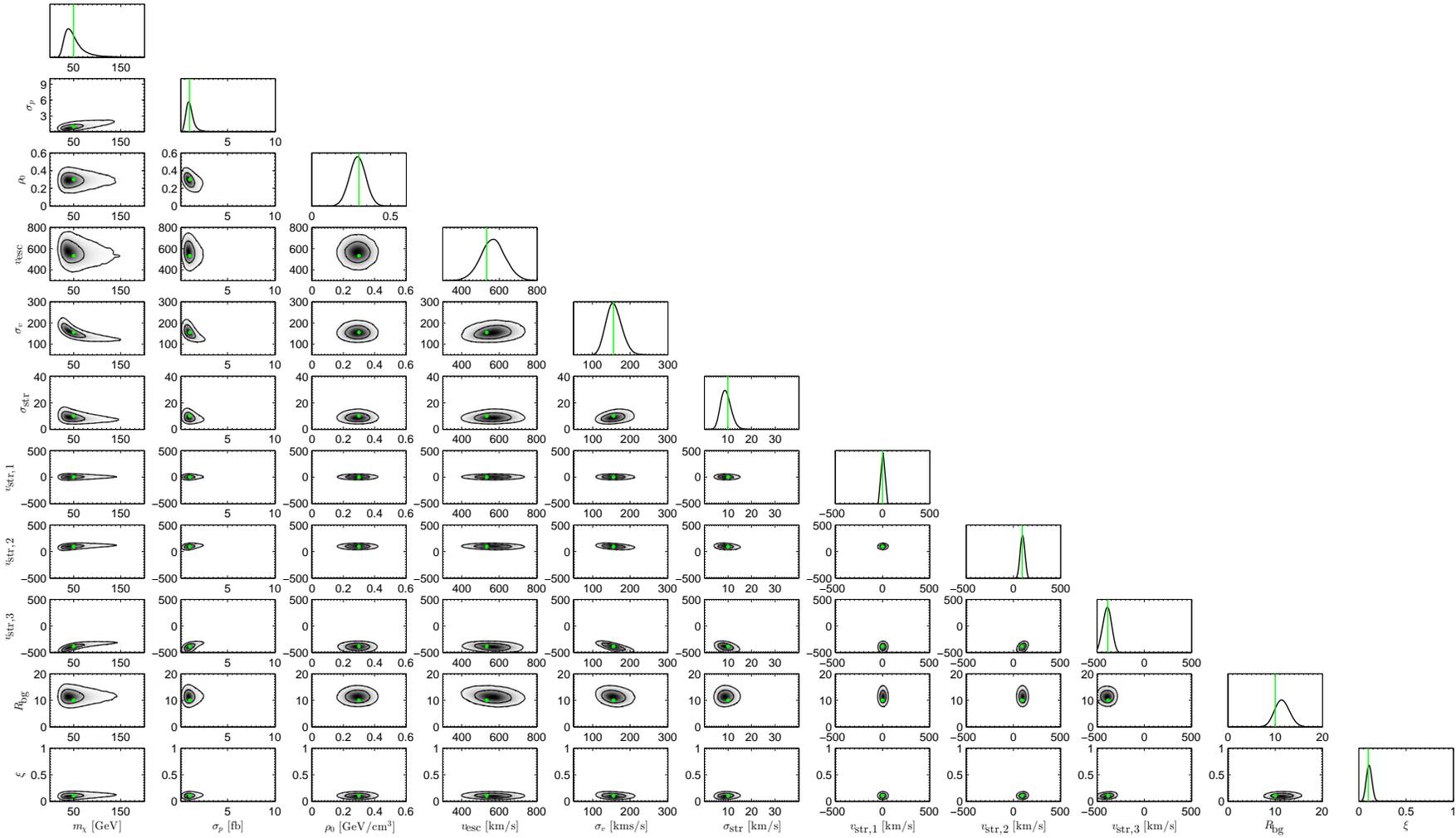}
			\caption{The 1 and 2-dimensional marginalised posterior probability distributions for the 11 parameters of the MB+stream model. The contours indicate the 68\% and 95\% credible regions. The green dots/lines indicate the location of the input parameters.}
			\label{posts_11params}
		\end{figure*}
		Whilst the questions that we are trying to address are frequentist in nature, there is much to be gained from parameter estimation by Bayesian inference. We can test this approach by reconstructing the input parameters used to generate a set of mock data, again initially studying the example case of a Sagittarius-like stream and a $m_{\chi}= 50 \, {\rm GeV}$ WIMP, using the same benchmark parameters and experimental configuration as in Sec.~\ref{sec:directionalstats}.

		We sample the likelihood using the nested sampling software \textsc{Multi}N\textsc{est}~\cite{Multinest1, Multinest2} using 5000 live points, an evidence tolerance  factor of 0.05 and sampling efficiency of 0.3. The posterior distribution is the probability distribution of the parameters $\boldsymbol{\theta}$ given a data set $d$ and is defined by Bayes' theorem,

		\begin{equation}
			p(\boldsymbol{\theta} | {d}) = \frac{\pi(\boldsymbol{\theta}) \mathcal{L}({d} | \boldsymbol{\theta})}{E({d})},
		\end{equation}
		where $\mathcal{L}$ is the likelihood function, i.e. the probability of the data given parameters $\boldsymbol{\theta}$, $\pi(\boldsymbol{\theta})$ is the prior distribution for the parameters and $E(d)$ is the evidence (practically, the normalisation factor for the posterior distribution). For each parameter we use a flat prior with ranges indicated by the axes of Fig.~\ref{posts_11params}. We define minimum credible regions which give contours of the parameter space encompassing 95\% or 68\% of the posterior distribution and indicate how well the parameters are reconstructed from the data.  Figure ~\ref{posts_11params} shows the marginalised 1 and 2-dimensional posterior probability distributions of the parameters. The marginalised probability distributions are calculated by integrating the distribution over the remaining parameters, e.g. for parameters $\boldsymbol{\theta} = \lbrace\theta_1, ..., \theta_n\rbrace$,
		\begin{equation}
			p(\theta_1,\theta_2 | d) = \int p(\boldsymbol{\theta} | d) \prod_{i=3}^n \textrm{d}\theta_i \, .
		\end{equation} 

		While the stream parameters are recovered well, the largest uncertainties are present in the halo and WIMP parameters, in particular $m_\chi$. The effect of the Gaussian parametrisation of $v_\textrm{esc}$ and $\rho_0$ is apparent and equivalent to using a Gaussian prior. There is a still some correlation in the $\rho_0-\sigma_p$ plane but the degeneracy is broken. The halo parameters are reconstructed less accurately when the bare likelihood function is used, however the Gaussian parameterisation is representative of the astrophysical uncertainties.   

		Our conclusions are not sensitive to the form assumed for the background energy spectrum. In Appendix A we show that the 
		the maximum likelihood estimators for the stream parameters for a Sagittarius-like stream with an exponential background are almost identical to those with a flat background.

		We have shown that good constraints can be placed on the parameters describing a stream, if the correct model is used in constructing the likelihood function for the data. Importantly, we note that the exposure times needed to make these constraints  $\sim 10$ kg yr are significantly shorter than are needed when the non-parametric directional tests are used. Our results agree with a similar work by Lee and Peter~\cite{LeePeter}. They likewise found that the constraints improved with the use of the full energy and direction information, however they used a different parameterisation, subsuming the degenerate local density and cross section into a single parameter, fixing the total number of WIMP and background events and normalising out the amplitude of the signal. Here, in order to achieve independence under changing halo, WIMP and experimental parameters, we quote exposure time and not total number of events and must parameterise the likelihood function with local density and cross section and also use the background rate $R_\textrm{bg}$ as an input parameter.

	\subsection{Likelihood ratio}

		\begin{figure}
			\centering
			\includegraphics[trim = 7mm 0mm 0mm 0mm, clip, width=0.45\textwidth]{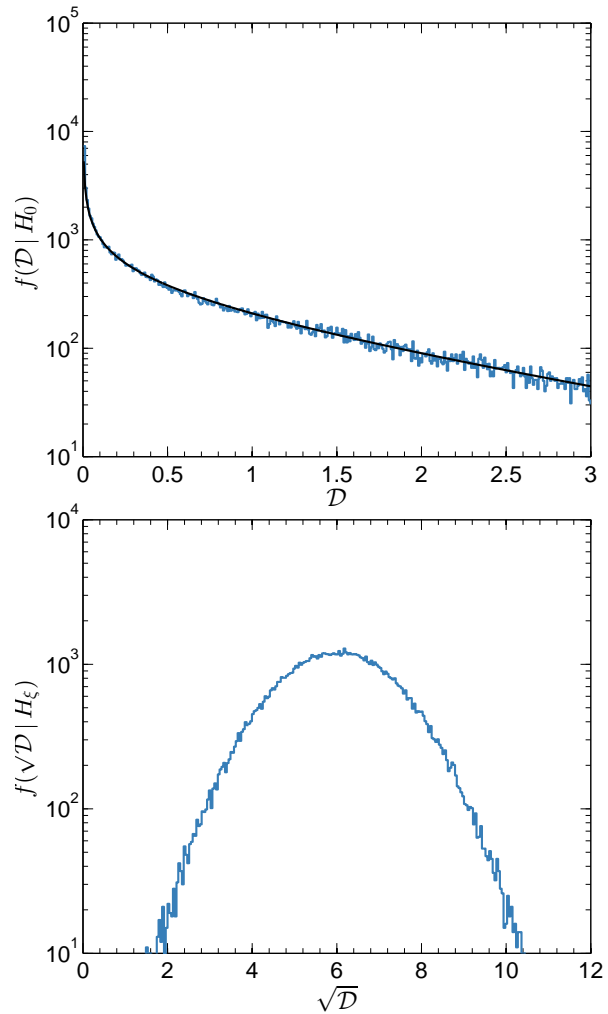}
			\caption{Distributions of the profile likelihood ratio test statistic, $f(\mathcal{D})$, under the null hypothesis of zero stream density, $\xi = 0$ (top) and the distribution of the significance in units of $\sigma$, $f(\sqrt{\mathcal{D}})$, under the alternative hypothesis with $\xi = 0.1$ (bottom) generated from $10^5$ Monte Carlo experiments. The black curve in the top panel shows a fitted $\chi_1^2$ distribution, demonstrating that Wilk's theorem holds. The distributions were generated from the input parameters in Tab. \ref{tab:partable} with an exposure of 10 kg yr and energy window of $[5,100]$ keV.}
			\label{llratio_null_alternative_dists}
		\end{figure}
		\begin{figure}
		  \centering
		  \includegraphics[trim = 0mm 0 0mm 0mm, clip, width=0.45\textwidth]{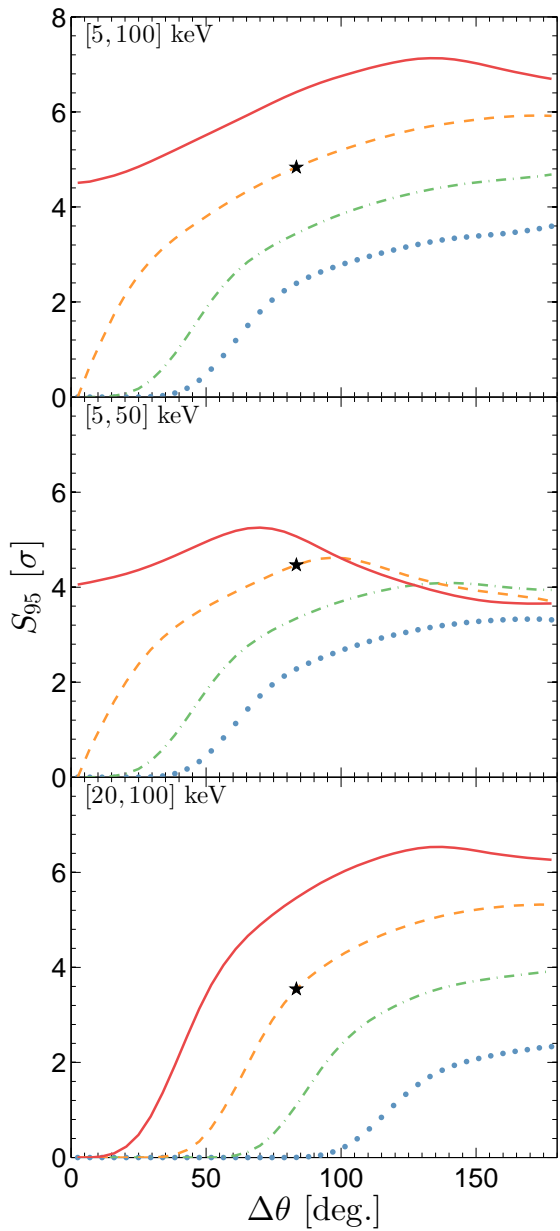}
		  \caption{The significance obtainable by 95\% of experiments, $S_{95}$,  in units of $\sigma$, as a function of the angle between the Solar and stream velocities, $\Delta \theta$, for a 50 GeV WIMP. The curves correspond, from top to bottom in each panel, to $v_\textrm{str} = 500$ km s$^{-1}$ (red solid line), 400 km s$^{-1}$ (orange dashed), 300 km s$^{-1}$ (green dot-dashed), 200 km s$^{-1}$ (blue dotted). The three panels are for energy windows  $[5,100]$ keV, $[5,50]$ keV and $[20,100]$ keV, from top to bottom. The star denotes the result for the Sagittarius stream.}
		  \label{S95_DeltaTheta}
		\end{figure}
		\begin{figure*}
		  \centering
		  \includegraphics[trim = 0mm 0 0mm 0mm, clip, width=0.9\textwidth]{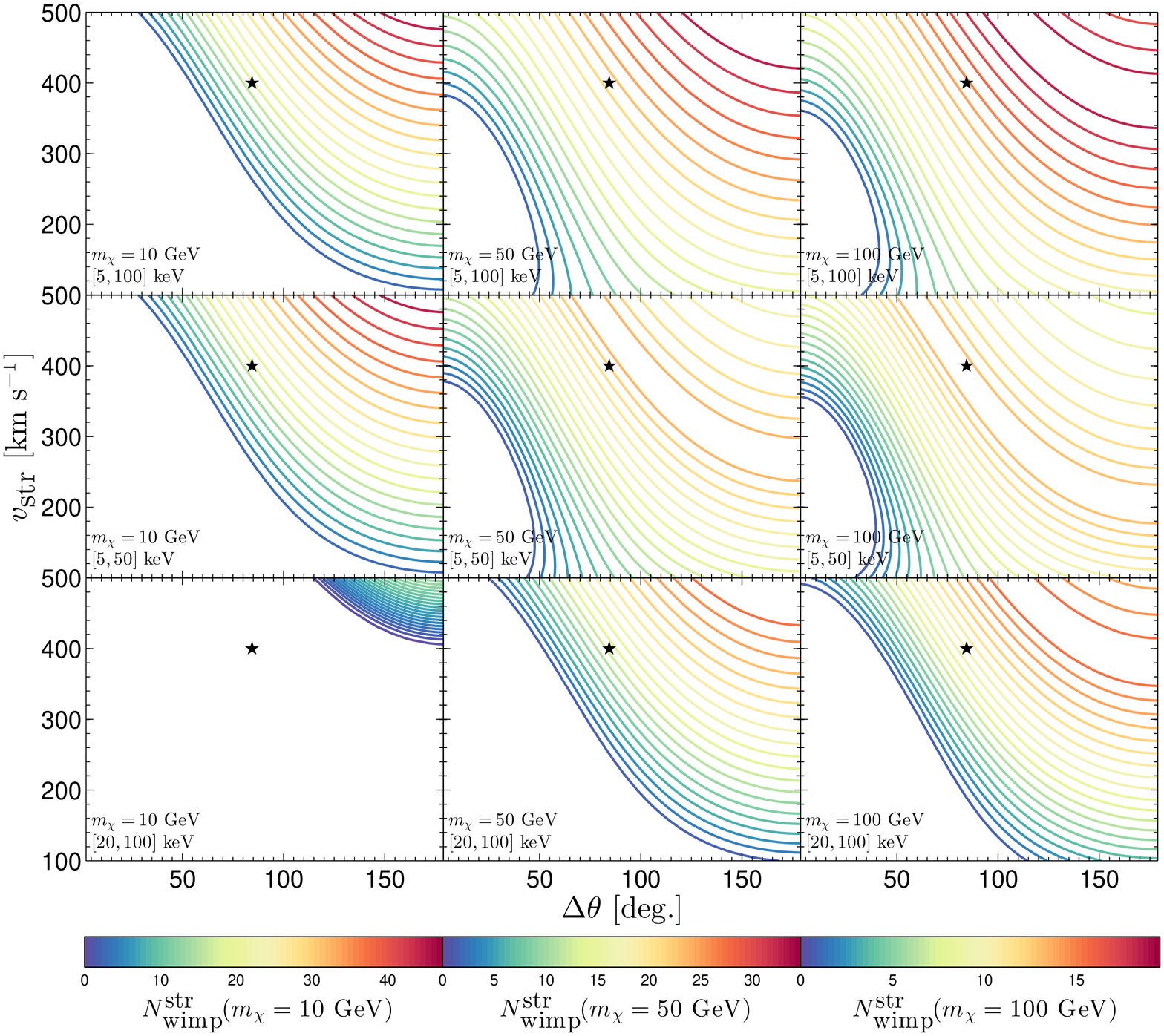}
		  \caption{The number of recoils from stream WIMPs, $N_\textrm{wimp}^\textrm{str}$, as a function of stream speed, $v_{\rm str}$ and direction, $\Delta \theta$, for $m_{\chi} =10, 50$ and $100 \, {\rm GeV}$ (from left to right) and energy windows $[5,100]$ keV, $[5,50]$ keV and $[20,100]$ keV (from top to bottom). The star denotes the speed and direction of the Sagittarius stream.}
		  \label{N_wimp_str_location_cont}
		\end{figure*}
		We now turn our attention towards the profile likelihood ratio test, a frequentist test that we will use to assess the detectability of streams with different velocities in a parametric way. Model comparison is performed by considering two hypotheses, a null hypothesis, $H_0$, that the WIMP distribution is described by a MB distribution alone with no stream component and an alternative hypothesis, $H_\xi$, where the density of the stream is non-zero. The profile likelihood ratio test utilises the fact that the model best describing the null hypothesis is contained within the model best describing the alternative hypothesis. We use this test as it is computationally easier than performing Bayesian inference over a range of input velocities. The likelihood ratio between the null and alternative hypotheses is
		\begin{equation}
			\Lambda = \frac{\mathcal{L}(\hat{\hat{\boldsymbol{\theta}}}, \xi = 0)}{\mathcal{L}(\hat{\boldsymbol{\theta}} )} \, ,
		\end{equation}
		where $\hat{\boldsymbol{\theta}}$ are the maximum likelihood estimators in the alternative model, and $\hat{\hat{\boldsymbol{\theta}}}$ are the maximum likelihood estimators evaluated when the stream density fraction is fixed at zero, $\xi = 0$. The likelihood ratio test statistic is then defined as
		\begin{equation}
			\mathcal{D} = \left\{ \begin{array}{rl}
			-2\ln \Lambda  & \, \, 0\le \hat{\xi}\le1 \,,\\
			0  & \, \, \hat{\xi}<0, \, \, \hat{\xi}>1 \,.
			\end{array} \right. 
		\end{equation}

		Next we require an expression for the statistical significance of a particular measured test statistic value. This requires knowledge of how the profile likelihood ratio test statistic $\mathcal{D}$ is distributed in the case that the null hypothesis is true, i.e. if the observed value is $D_\textrm{obs}$:

		\begin{equation}
			S = \int_0^{\mathcal{D}_\textrm{obs}} f(\mathcal{D} | H_0) \, \textrm{d}\mathcal{D} \,.
		\end{equation}

		It is known however from Wilk's theorem~\cite{Wilks} that the distribution of the profile likelihood ratio test statistic in the case that the null hypothesis is true asymptotes towards a $\chi_1^2$ distribution. Hence the discovery significance is defined as $S = \textrm{erf}\left(\sqrt{\mathcal{D}_\textrm{obs}/2}\right)$. However as we will be quoting quite high values of significance it is simpler to write them in units of standard deviation, $\sigma$, i.e. $S = \sqrt{\mathcal{D}_\textrm{obs}}$.  So a value of $\mathcal{D}_\textrm{obs} = 1$ corresponds to a 1$\sigma$ result or a significance of 68\%. The significance obtainable by 95\% of mock experiments, $S_{95}$, is then found by solving the equation,
		\begin{equation}
		 \int_0^{S_{95}} f(\sqrt{\mathcal{D}})\textrm{d}\sqrt{\mathcal{D}} = 0.95 \, .
		\end{equation}
		In Fig.~\ref{llratio_null_alternative_dists} we show the distributions of $\mathcal{D}$ from $10^5$ mock experiments under the null and alternative hypotheses, for the Sagittarius-like stream given in Tab. \ref{tab:partable}.  For a non-zero stream density fraction the test statistic has a Gaussian distribution which moves further from $\mathcal{D}=0$ as the stream density is increased. Again, our conclusions here are not sensitive to the form assumed for the background energy spectrum. In Appendix A we show that the distribution of the likelihood ratio test statistic for a Sagittarius-like stream with an exponential background is almost identical to that with a flat background.

		As in Sec.~\ref{sec:directionalstats} we now study the performance of the test over a range of stream velocities where we will again use the stream speed, $v_\textrm{str}$, and the angle between the Solar and stream velocities, $\Delta \theta$, to describe the stream velocity. In Fig.~\ref{S95_DeltaTheta} we plot the significance, in units of $\sigma$, obtainable by 95\% of hypothetical experiments, $S_{95}$, using the profile likelihood ratio test as a function of stream speed, $v_\textrm{str}$, and direction given by $\Delta \theta$, for a 50 GeV WIMP and the three energy windows considered previously. Figure \ref{S95_DeltaTheta}, and also Fig.~\ref{N_wimp_str_location_cont}, was generated using exposure times of 5 kg yr and the parameters not plotted were all taken to have the benchmark values given in Tab. \ref{tab:partable}. The test, by virtue of being parametric, performs much better than the non-parametric tests. For the example of the Sagittarius stream, indicated in the Figures by a star, the tests for the three energy windows detect the stream at $4 - 5 \sigma$ in 95\% of experiments. For the same exposure, the non-parametric tests could only reach a value of $S_\textrm{95}$ between 0.1 and 0.2.

		The enhancement in performance is also due to the use of the full energy and direction data whereas before only the direction information was used. Furthermore the tests achieve high significance over a wide range of stream velocities, with the limiting factor being the number of WIMPs coming from the stream, $N_\textrm{wimp}^{\textrm{str}}$, as can be seen by comparing the two Figures. For low values of $\Delta\theta$, where the number of stream WIMPs drops to zero, the significance can be seen to do likewise. There is similarly a dependence on the energy window of the detector which causes a reduction in the number of stream WIMPs when the stream becomes boosted past the maximum of the energy window. This can be seen clearly in the $[5,50]$ keV case. For a stream speed of 500 km s$^{-1}$, the significance begins to decrease for $\Delta\theta > 70^\circ$ and drops by $1.6\sigma$ up to $\Delta \theta = 180^\circ$. However the significance for faster stream speeds is enhanced over what might be expected simply from the dependence on $N_\textrm{wimp}^\textrm{str}$. This is due to faster streams becoming more prominent because of the exponential drop off with energy of the event rate for the smooth halo.

		The number of events from WIMPs in the stream, $N_\textrm{wimp}^\textrm{str}$, plays a key role in determining the detectability of a stream. Therefore in Fig.~\ref{N_wimp_str_location_cont} we show $N_\textrm{wimp}^\textrm{str}$  as a function of the stream speed and direction for the three energy windows and $m_{\chi} =10, 50$ and $100$ GeV. As discussed earlier, provided the WIMP mass is large enough that the stream recoils are above the energy threshold, then increasing the WIMP mass decreases the number of stream events. The energies of the stream recoils depend on the stream velocity as well as the WIMP mass. Increasing the threshold energy to 20 keV or reducing the maximum energy to 50 keV reduces the size of the region of steam speed-angle parameter space within which the number of stream events is large enough for the stream to be detected. For a 10 GeV WIMP and a 20 keV threshold a stream will only be detectable if it has a high speed and a large angle relative to the lab velocity.

\section{\label{sec:summary}Summary}
	Using both non-parametric directional tests and a profile likelihood test, we have shown that there are reasonable prospects for the detection of a moderately high density tidal stream by a future directional detector. We have looked at the concrete example of a Sagittarius-like stream and also explored the dependence on the parameters of the stream, namely its speed, direction, dispersion and density. 
	
	Using non-parametric directional statistics the detection of a Sagittarius-like stream would need a total  of around 900 events, but using Bayesian parameter estimation good constraints can be placed on the stream parameters with around 300 events, independent of astrophysical uncertainties. For WIMP-nucleon spin-dependent cross sections just below the current exclusion limits~\cite{Xenon} this corresponds to around 10 kg yr exposure, which is within the scope of future directional detectors~\cite{Ahlen}.
	
	We have shown that the detectability of a stream is highly dependent on its speed and position in relation to the lab velocity. Faster streams that are oriented at $90^\circ$ with respect to the lab velocity are most easily detected when using directional statistics as the signal in this case deviates furthest from the smooth halo signal. Using a profile likelihood ratio test, there is weaker dependence on the position of the stream as the likelihood does not favour a particular directionality of the signal, but there is naturally a dependence on the the number of WIMPs originating from the stream, which indirectly leads to a dependence on stream direction. Faster streams can be detected with higher significance, as they cause more events and are also more prominent in the energy spectrum, due to the exponential drop off with energy of events from the smooth halo. A similar conclusion is likely for parameter estimation as weaker constraints would be made on parameters of the stream with fewer WIMPs from the stream. The estimation of halo and WIMP parameters would also be dependent on the stream velocity, if the stream is boosted out of the energy window of the detector there would be fewer events overall which would weaken the relative constraints. 

	The advantage of using non-parametric tests is that one need not assume a model to describe the data, simply that the data satisfy either a null or alternative hypothesis. However non-parametric tests will always return a less significant result than parametric tests. The likelihood analyses also make use of both the energy and direction information of the recoils whereas the non-parametric tests are direction only. The Kuiper and Median direction tests may be improved with the use of energy information however this would result in a greater dependence on the WIMP mass and would be expected to perform poorly with low numbers of recoils.

	We have also investigated a number of experimental considerations. Firstly, as might be expected, a larger experimental background rate results in poorer performance of the statistical tests, but the dependence on the background rate flattens at very large values. A crucial factor is the energy window of the detector, where for the detection of streams the window must cover the range of recoil energies from stream WIMPs. This is dependent on both the speed of the stream and its direction in relation to the lab velocity as we have demonstrated. 

	The statistical methods we have studied here may all be applied in a practical setting on real data, however there remain some experimental issues that may affect the detectability of streams. The mock detector we have considered here has perfect angular resolution. In practice a finite angular resolution of around $10^\circ$ is thought to be achievable for future detectors \cite{MIMAC}. This  will have some impact on the power of an experiment to detect fine substructure, however the results here are likely to remain valid if finite angular resolution is taken into account either through a binned analysis or by adding in an uncertainty in the recoil data. Another experimental concern is that of sense recognition (i.e. distinguishing between $\hat{\textbf{q}}$ and $-\hat{\textbf{q}}$) which remains an issue for some current directional detectors. A lack of sense recognition has been shown to increase the number of events needed to reject an isotropic non-WIMP background in the initial detection phase~\cite{Green3} and is likely to be problematic for the detection of substructure and new tools may be needed to reconstruct the velocity distribution if it is still an issue in the future. 

	To conclude, we have shown that if a moderately high density WIMP stream is present in the Solar neighbourhood then there are good prospects for its detection using directional dark matter detectors.

\begin{acknowledgments}
	The authors thank Mattia Fornasa and Tasos Avgoustidis for useful discussions. CAJO and AMG are both supported by the STFC.
\end{acknowledgments}

	\section*{Appendix A: Background model dependence}
	In this Appendix we studying the effects of assuming an exponential, rather than flat, background energy spectrum on the stream parameter estimation in Sec.~\ref{param-reconstruct}. The ``worst possible'' energy dependence would be an exponential with a characteristic scale that is similar to that of the WIMP signal i.e.
		\begin{equation}\label{expback}
		 \frac{\textrm{d}^2 R}{\textrm{d}E\textrm{d}\Omega_q} = \frac{R_\textrm{bg}}{4 \pi E_\textrm{bg}} \frac{e^{-E/E_\textrm{bg}}}{e^{-5\, \textrm{keV}/E_\textrm{bg}} - e^{-100\, \textrm{keV}/E_\textrm{bg}}},
		\end{equation}
		where $R_\textrm{bg}$ is defined as the rate observed in an experiment with a window of $[5,100]$ keV and $E_\textrm{bg} = 17.5 \, {\rm keV}$ for a background which mimics the energy dependence of a $50 \, {\rm GeV}$ WIMP. The previous 11-dimensional parameter space is expanded to 12-dimensions with the addition of $E_\textrm{bg}$.
		  \begin{figure*}
			  \centering
			  \includegraphics[trim = 0mm 0 0mm 0mm, clip, width=0.9\textwidth]{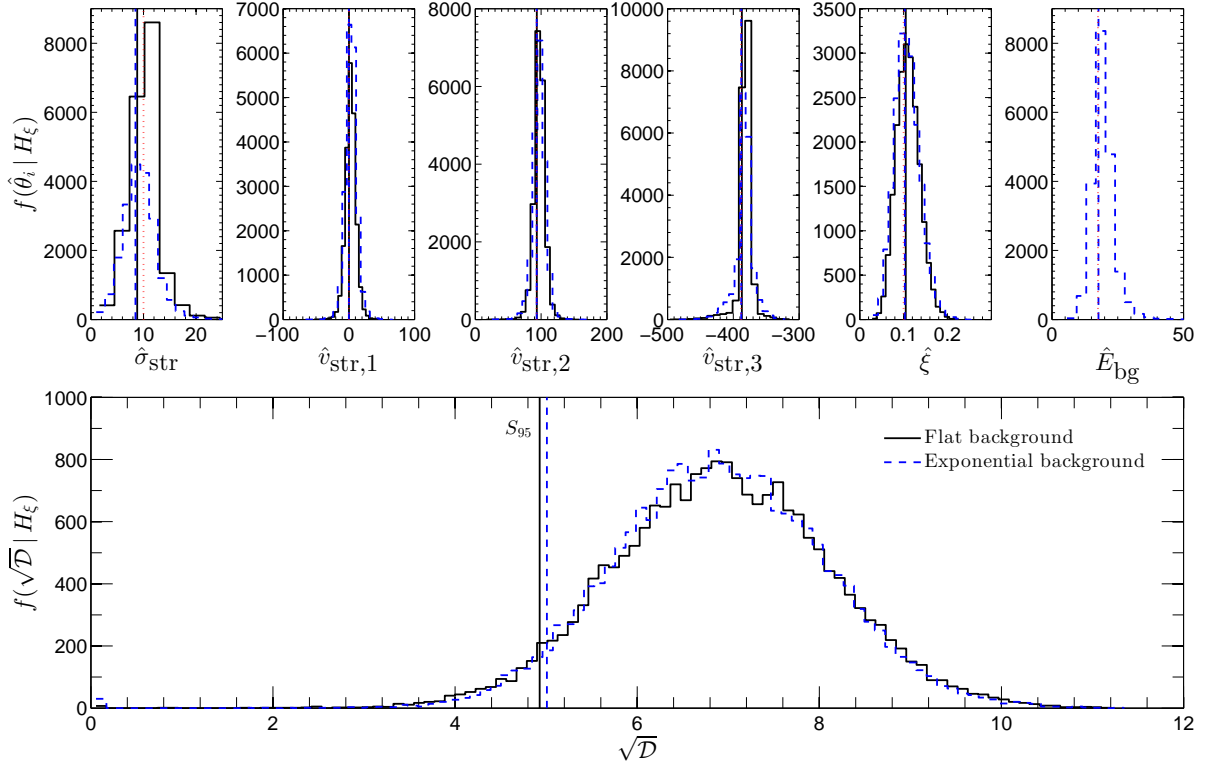}
			  \caption{Top row:
			  The distributions of the maximum likelihood estimators (MLE) of the stream parameters and the background energy dependence parameter $E_\textrm{bg}$ obtained over 20000 Monte-Carlo runs for the flat and exponential background models (solid and dashed lines respectively). The vertical lines indicate the mean MLE values and the red dotted vertical lines the input parameter values. Bottom plot: the distribution of the likelihood ratio test statistic for the flat and exponential background models (solid and dashed lines respectively), the vertical lines mark the location of the significance achievable in 95\% of experiments, $S_{95}$.}
			  \label{backgroundtest}
		\end{figure*}

		In Fig. \ref{backgroundtest} we plot the distribution of the likelihood ratio test statistic and the maximum likelihood estimators for the stream parameters of a Sagittarius-like stream for both flat and exponential backgrounds. It can be seen that the the distribution of the test statistic and the accuracy with which the parameters are recovered are nearly identical for the two background models.

\end{document}